\documentclass[allclo]{FBSart}
\usepackage{amsfonts}
\usepackage{amssymb}
\usepackage{epsfig}

\title{Two-nucleon problem in semi-relativistic baryon chiral perturbation theory}
\author{D.~Djukanovic, J.~Gegelia\thanks{\textit{Alternative address:}
Institute of High Energy Physics, Tbilisi State University, 0186
Tbilisi, Georgia}, S.~Scherer, M.~R.~Schindler } \institute{Institut
f\"ur Kernphysik, Johannes Gutenberg-Universit\"at, D-55099 Mainz,
Germany}

\runningauthor{D.\,Djukanovic et al} \runningtitle{Style File for
Few-Body Systems} 

\begin{document}

\maketitle
\begin{abstract}
We consider a symmetry-preserving approach to the nucleon-nucleon
scattering problem in the framework of the higher-derivative
formulation of baryon chiral perturbation theory. Within this
framework the leading-order amplitude is calculated by solving
renormalizable equations and corrections are taken into account
perturbatively.

\end{abstract}

    It has been claimed that Weinberg's program for the few-nucleon sector of
baryon chiral perturbation theory \cite{Weinberg:rz} encounters
conceptual problems. The NN potential of the effective field theory
(EFT) is non-renormalizable already at leading order (LO). The
iteration of the potential in the LS equation generates divergent
terms with structures which are not included in the original
potential. Therefore, the renormalization of the solution to the LS
equation requires the contributions of an infinite number of
higher-order counter-terms.
   It has been argued that the coefficients of the divergent parts of the
counter-terms contributing in low-order calculations would set the
scale of the corresponding renormalized couplings
\cite{Kaplan:1998tg}.
   Therefore, even if these couplings were natural at some value
of the renormalization scale, they would become unnaturally large
for slightly different values of this parameter.
   This problem, in different variations,
has been addressed as the inconsistency of Weinberg's approach. On
the other hand, cutoff EFT advocates the point of view that the
''consistency problem'' of Weinberg's approach is irrelevant and one
can perform the calculations by suitably choosing the cutoff
parameter (see, e.g.,
\cite{Epelbaum:2005pn,Lepage:1997cs,Gegelia:gn,Gegelia:1998iu,Gegelia:2004pz}).
The controversy still remain unresolved.

Here, we shortly present a new approach developed in
Ref.~\cite{Djukanovic:2006mc}.

The standard effective Lagrangian leads to UV divergences. These can
be handled by considering a nucleon propagator with an improved UV
behavior \cite{Djukanovic:2004px}
\begin{equation}
i\,S_F(p)=\frac{i}{p \hspace{-.45em}/\hspace{.1em}
-m+i\,\epsilon}\,F(L^2,\vec p\,{}^2)\,, \label{fp}
\end{equation}
where the regulating function $F$ can be chosen as e.g.
\begin{equation}
F(L^2,\vec p\,{}^2)=\frac{L^{4\,N_\Psi}}{\left( L^2+\vec
p^2\right)^{2\,N_\Psi}}\, \ {\rm or} \ F(L^2,\vec p\,{}^2)=\exp
\left\{ -2 \left( \frac{\vec p\,^2 }{L^2}\right)^{N_\Psi}\right\}\,.
\label{ff}
\end{equation}

The equation for the nucleon-nucleon scattering amplitude reads
\begin{equation}
T = V + V \,G \ T \,.\label{EqForGF}
\end{equation}
Expanding $T$, $V$, and $G$ in small parameters (like the pion mass
and small momenta),
\begin{equation}
G =  G_0 +G_1 + \cdots\,,  \ V = V_0 +V_1 + \cdots\,, \ T = T_0 +T_1
+\cdots\,, \label{Texpanded}
\end{equation}
and substituting in Eq.~(\ref{EqForGF}), we solve $T$ order by order
(analogously to the KSW approach \cite{Kaplan:1998tg}). At leading
order we obtain
\begin{equation}
T_0 = V_0 + V_0\, G_0\, T_0\,. \label{LOgeq}
\end{equation}
Here, the LO two-nucleon propagator $G_0=-i\,S_F^{(1)} S_F^{(1)}$
and $S_F^{(1)}$ is defined by
\begin{eqnarray}
S_F(p) &=&\frac{m\, F(L^2,\vec p\,{}^2)\, (1+\gamma^0)/2
}{\sqrt{\vec p^2+m^2}\,\left( p_0-\sqrt{\vec p^2+m^2}+
+i\,\epsilon\right)}+\cdots = S_F^{(1)}(p)+\cdots
\,.\label{Sfpexpanded}
\end{eqnarray}

Using the LO amplitude $T_0$, the NLO amplitude $T_1$ is calculated.
Analogously, $T_0$ and $T_1$ can be used to calculate the NNLO
amplitude $T_2$ etc.

\begin{figure}
\epsfig{file=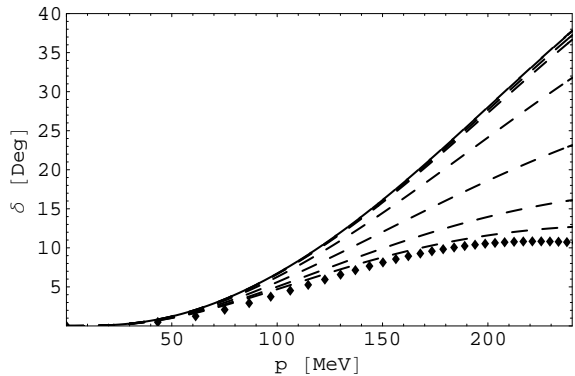,width=0.6\textwidth}
\caption[]{\label{nn3p0:fig} $^3P_0$ partial wave $np$ phase shifts
for different values of $L$ parameter compared with the data points
from the Nijmegen PWA \cite{Stoks:1993tb}. The curves are shown for
$L=$ 750, 1000, 2000, 6000, 20000, 30000, 100000, 500000 (MeV), the
solid line corresponding to the largest value.}
\end{figure}

    The LO equation in the center-of-mass frame reduces to
\begin{eqnarray}
{\cal T}_0\left( \vec p\,',\vec p\right)= {\cal V}_0\left( \vec
p\,',\vec p\right)- m\, \int \frac{d^3 k}{(2\,\pi)^3} \ {\cal
V}_0\left( \vec p\,',-\vec k\right) \, \frac{1}{ \vec p^2-\vec
k^2+i\,\epsilon} \,{\cal T}_0\left( -\vec k,\vec p\right)\,,
\label{MeqLOk0integratedFinal}
\end{eqnarray}
where
\begin{eqnarray}
{\cal T}_{0}\left( \vec p\,',\vec p\right) &=& \frac{\tilde
T_{0}\left( \vec p\,',\vec p\right)\, F(L^2,\vec p\,'\,{}^2)\,
F(L^2,\vec p\,{}^2)
\,m^2}{[(m^2+\vec p\,'\,^2)(m^2+\vec p\,^2)]^{1/2}}\nonumber\\
{\cal V}_{0}\left( \vec p\,',\vec p\right) &=& \frac{\tilde
V_{0}\left( \vec p\,',\vec p\right)\,F(L^2,\vec p\,'\,{}^2)\,
F(L^2,\vec p\,{}^2)\,m^2}{[(m^2+\vec p\,'\,^2)(m^2+\vec
p\,^2)]^{1/2}}\,. \label{newcaldefinitions}
\end{eqnarray}
Here, $\tilde V_0$  is the sum of the standard contact interaction
($V_{0,C}$) and one-pion exchange ($\tilde V_{0,\pi}$) parts of the
LO potential \cite{Epelbaum:2005pn}. The potential ${\cal V}_0(\vec
p\,',\vec p)$ has a milder ultraviolet behavior than $\tilde
V_{0}\left( \vec p\,',\vec p\right)$. In the limit $L\to\infty$ the
OPE part of the effective potential ${\cal V}_0(\vec p\,',\vec p)$
generates no divergences in the $^1S_0$ wave. The $^3S_1$ wave
requires a single counter-term, which is present in our LO
potential. For higher partial waves the singular behavior of the
potential is screened by the angular momentum barrier and therefore
no counter-terms are required. The numerical analysis confirms the
above conclusions. The $^3P_0$ phase shifts calculated for different
values of the parameter $L$ are shown in Fig.~\ref{nn3p0:fig}.
Different choices of $L$ correspond to different renormalization
schemes \cite{Djukanovic:2006mc}. By choosing the renormalization
scheme properly one can improve the convergence of perturbative
series for physical quantities. The best description of the phase
shifts at leading order is obtained for $L\sim 750$ MeV.
Corrections to any finite order can be expressed as a sum of a
finite number of Feynman diagrams where $T_0$ is also interpreted as
NN vertex. The UV behavior of $T_0$ guarantees that all divergences
generated in the limit $L\to\infty$ in corrections at any finite
order can be canceled by counter-term contributions also present at
the given order. As a result our new approach is free of
''consistency problems''.

\begin{acknowledge}
D.D., J.G. and M.R.S.~acknowledge the support of the Deutsche
Forschungsgemeinschaft (SFB 443).
\end{acknowledge}

\end{document}